\documentclass{iopconfser}

\usepackage{xcolor}
\usepackage{mathtools,amssymb}

\begin{document}

\title{Singularity resolution in the backreacted Schwarzschild geometry from 2D
matter with negative central charge\footnote{Expanded version of the talk given by F.J. M-G at the 24th International Conference on General Relativity and Gravitation and the 16th Edoardo Amaldi Conference on Gravitational Waves, to appear in the conference proceedings.}}

\author{F J Marañón-González$^{1}$, A del Río$^{2}$ and J Navarro-Salas$^{1}$}

\affil{$^1$Instituto de Física Corpuscular (IFIC), CSIC-Universitat de València and Departament de Física Teòrica, UV, Spain. \newline}

\affil{$^2$Departamento de Matem\'aticas, Universidad Carlos III de Madrid,  Avda de la Universidad 30, 28911 Legan\'es, Madrid, Spain.}

\email{jmarag@ific.uv.es}

\begin{abstract}

Two-dimensional dilaton gravity provides a valuable framework to study the dynamics of quantum black holes. These models are often coupled to conformal scalar fields, which capture essential quantum effects such as the trace anomaly, while remaining analytically tractable. From the viewpoint of two-dimensional quantum field theory, unitary theories require a positive central charge. However, theories with a total negative central charge naturally arise from the contribution of the Faddeev-Popov ghosts to the effective action.   Recent analyses of the Callan-Giddins-Harvey-Strominger (CGHS) model with a Russo-Susskin-Thorlacius (RST) counterterm have shown that a negative central charge can remove curvature singularities in the backreacted geometry. In this work, we argue that singularity resolution arises from the negative central charge itself, rather than the particular dynamics of a given model. To support this, we present analogous results in spherically reduced Einstein gravity.

\end{abstract}

\section{Introduction}

The study of quantum effects in a gravitational setting quickly becomes a highly intricate subject. On the gravitational side, General Relativity (GR) is known to be non-renormalizable, which signals the need for an improved framework to consistently address quantum gravity. Proposals such as String Theory and Loop Quantum Gravity fall into this category, each with their own intrinsic challenges. An alternative route is to introduce quantum effects only through the matter sector, while treating spacetime itself with a classical gravitational theory. This line of research, known as Quantum Field Theory (QFT) in Curved Spacetime \cite{birrell-davies}, is arguably the most conservative and well-grounded approach given our current knowledge. In this work, we adopt this perspective and argue that certain unconventional quantum field theories can naturally resolve the Schwarzschild singularity, leading to semiclassical geometries  that coincide with the black hole metric at asymptotic infinity, but without horizons and with  curvature bounded everywhere.

When formulating a QFT on a curved background, non-linear matter–gravity interactions generate new divergences, making renormalization more involved. This renormalization is background-dependent, so the spacetime geometry must be specified to compute quantities such as the components of the renormalized stress–energy tensor. While often a good approximation, there are points or even entire regions of classical spacetime where the renormalized stress–energy tensor diverges. In such cases, quantum backreaction can no longer be ignored, and the spacetime geometry is expected to change significantly. For instance, in the case of a scalar field on a Schwarzschild background with the static Boulware vacuum, the stress–energy tensor diverges at the event horizon \cite{Christensen:1977jc, Candelas:1980zt}.
It is therefore natural to investigate scenarios in which quantum backreaction becomes strong enough to eliminate the event horizon, or possibly even resolve the central singularity.

Our work was initially motivated by earlier studies on singularity resolution in two-dimensional gravity, particularly in the CGHS model \cite{PSS0, PSS1, PSS2}. This contribution presents our recent extension of the approach in Refs.~\cite{PSS0, PSS1, PSS2} to spherically reduced four-dimensional GR \cite{nosotros}. We also offer a prospective discussion on the connection between negative central charge and the mechanism of singularity resolution.

\section{Effective semiclassical theory}
Two-dimensional effective theories provide a convenient framework in which backreaction can be incorporated dynamically into the geometry. They gained prominence through the Jackiw–Teitelboim \cite{J, T} and Callan–Giddings–Harvey–Strominger (CGHS) \cite{CGHS} models. An additional advantage of these theories is that, unlike in most higher-dimensional cases, the semiclassical field equations remain second order. The matter sector is typically taken to consist of $N$ minimally coupled scalar fields $f_i$, which are conformal in two dimensions,
\begin{equation}
    S_m= -\frac{1}{2}\sum_{i=1}^N\int d^2x \sqrt{-g}\, \nabla_a f_i \nabla^a f_i\ .
\end{equation}
Such a theory has a vanishing trace classically. But once it is renormalized in a curved background, a non-vanishing trace anomaly arises: 
\begin{equation}
    g^{ab}\langle T_{ab}\rangle = \frac{\hbar C}{24\pi} R \ .
\end{equation}
Here, $R$ is  the Ricci scalar of the two-dimensional spacetime and $C$ is the so-called central charge $C$ of the theory. For a set of $N$  scalar fields, $C=N$. The trace anomaly is independent of the underlying quantum state of the matter sector. 
Interestingly, the trace condition, together with $\nabla^a\langle T_{ab} \rangle = 0$, are enough to determine $\langle T_{ab} \rangle$ in a conformal gauge $ds^2=-e^{2\rho}dx^+ dx^-$. Specifically, one gets
\begin{equation}
\begin{aligned}
    \langle T_{+-} \rangle &= - \frac{\hbar C}{12 \pi}\partial_+\partial_-\rho \\ 
    \langle T_{\pm\pm} \rangle &= - \frac{\hbar C}{12\pi} \left[ (\partial_\pm\rho)^2 - \partial_\pm^2 \rho + t_\pm(x^\pm)  \right] \ ,
\end{aligned}
\end{equation}
where the chiral functions $t_\pm$, fixed by assuming suitable boundary conditions, characterize the vacuum state. Equivalently, that stress-energy tensor can also be derived from the Polyakov (non-local) action \cite{P},
\begin{equation}
    S_P = -\frac{\hbar C}{96\pi} \int d^2 x \sqrt{-g} \, R \, \Box^{-1}R \ .
\end{equation}
For an introduction to these aspects of two-dimensional semiclassical gravity in the context of the Jackiw–Teitelboim and CGHS models, see \cite{FN05}.

Regarding the gravitational sector, we choose the four-dimensional Hilbert-Einstein action of GR, and construct an effective two-dimensional theory by imposing first spherical symmetry of the solution, and then integrating out  the angular variables. By doing this we get a dilatonic two-dimensional theory of gravity \cite{FN}
\begin{equation}
    S_{cl} = \frac{1}{2 G_2} \int d^2 x \sqrt{-g} \, e^{-2\phi} \left[ \frac{R}{2} + (\nabla \phi)^2 + \frac{e^{2\phi}}{r_0^2} \right] \ ,
\end{equation}
where the dilaton field $\phi$ is related to the four-dimensional radial coordinate $r^2 = r_0^2\,  e^{-2\phi}$, which becomes a dynamical function of the two-dimensional coordinates $x^\pm$. On the other hand, $G_2=G/r_0^2$ is the effective coupling constant, and $r_0$ is an arbitrary length scale.

The complete semiclassical theory, including backreaction and vacuum polarization effects, is obtained from $S=S_{cl} + S_P + S_m$. The presence of $S_m$ allows for additional classical matter (which can be useful in dynamical scenarios such as semiclassical gravitational collapse). We focus now on static solutions in 
a semiclassical vacuum: $S_m\to 0$, but $S_P\neq0$.
Therefore, these solutions will depend only on the spacelike coordinate $x=(x^+-x^-)/2$, in which the static vacuum defined as $t_\pm=0$ is the Boulware vacuum. The field equations obtained from variation of the action become
\begin{eqnarray}
\partial_x^2\phi_{}-(\partial_x\phi)^2-2\partial_x\rho \partial_x\phi&=&\frac{C\lambda}{r_0^2} e^{2\phi}\left(\partial_x^2\rho_{}-(\partial_x\rho)^2\right)
\label{eq:cg-i}\ , \ \ \\
\partial_x^2\phi_{}-2(\partial_x\phi)^2+\frac{e^{2(\phi+\rho )}}{r_0^2}&=& \frac{C\lambda}{r_0^2} e^{2\phi}\partial_x^2\rho_{}
\label{eq:cg-ii}\ , \\
\partial_x^2\phi_{}-(\partial_x\phi)^2-\partial_x^2\rho_{}&=&0 \label{eq:cg-iii}
 \ , \end{eqnarray} 
 with the definition $\lambda\equiv\hbar G/(12\pi)$. One can impose a planar Schwarzschild black hole (i.e. the $t-r$ sector of the 4D solution) as an asymptotic solution as $x\to\infty$ by ensuring
\begin{equation}
\begin{split}
    ds^2 &\sim \left(1-\frac{2GM}{r}\right)(-dt^2+dx^2) \\ 
    x &\sim r + 2 GM  \log\left(\frac{r}{2GM }-1\right) \ ,
\end{split}
\end{equation}
since this is the exact solution of the classical ($\lambda=0$) field equations. Therefore, $x^{\pm}=t\pm x$ can be identified asymptotically with the  Eddington-Finkelstein coordinates of the classical solution. The semiclassical equations need to be solved numerically.

\subsection{Effect of the central charge sign}
If $C>0$, the backreaction of the quantum fields is strong enough to remove the horizon, but not enough to achieve singularity resolution: a null naked curvature singularity remains at a finite affine parameter distance inside $r=2M$. The geometry is that of an asymmetric singular wormhole, with the singularity truncating one throat and the other asymptotic to the planar Schwarzschild solution \cite{FFNOS06, Arrechea:2019jgx}. Recent works using approximation techniques in  four-dimensions  also show a similar behaviour \cite{pau, Arrechea:2023oax}.

The case $C<0$ has been analysed very recently \cite{nosotros}.  The semiclassical geometry retains the horizonless and asymptotic properties, but furthermore becomes nonsingular in the region $r<2M$. The Ricci curvature remains bounded at any point. It is important to note the non-perturbative nature of the solution, since $R\propto1/\hbar$ near the origin.

\section{A few comments on $C<0$}

This regime has not been widely explored, since two-dimensional conformal field theories typically require a positive central charge in order to preserve unitarity.
Nevertheless, there exist four-dimensional scenarios of physical relevance that motivate the consideration of negative central charges. In particular, recent interest has focused on conformally invariant scalar fields of vanishing (mass) dimension,
\begin{equation}
S= -\frac{1}{2} \int d^4x \sqrt{-g} \ \xi \triangle_4 \ \xi \ ,
\end{equation}
which can also yield negative central charges \cite{Gusynin89}. Such contributions may offer potential advantages both in the Standard Model \cite{BT,MVZ} and in gravity. 

Even in the two-dimensional setting \cite{CGHS}, there exists a natural rationale for assuming a negative central charge \cite{Strominger:1992zf}. This occurs when the total central charge (matter plus gravity) is dominated by the Faddeev–Popov ghost sector, which contributes $C=-26$. If the number $N$ of fields with positive central charge is sufficiently small, the overall central charge becomes negative \cite{Strominger:1992zf}, and singularity resolution may then occur.

One might argue that the resolution of singularities depends more on the details of the model than on the mere sign of the central charge. However, previous studies within the CGHS framework \cite{PSS0, PSS1, PSS2} have obtained the same qualitative behaviour. The model considered in \cite{PSS0, PSS1, PSS2} involved the same matter sector as here, but with the CGHS gravitational action replacing our gravitational term, together with the Russo–Susskind–Thorlacius (RST) counterterm \cite{Russo:1992ax}, which renders the model analytically tractable. Importantly, there are indications that the appearance of singularity resolution is not solely due to the presence of the RST counterterm. For example, employing alternative counterterms, such as those proposed in \cite{Cruz:1995zt}, one still finds that the semiclassical backreacted geometry remains free of curvature singularities when the central charge is negative.

These results suggest a certain universality in the connection between negative central charge and singularity resolution in two-dimensional gravity, and hint at the possibility of an analogous mechanism in four dimensions.

\section{Conclusions}

Negative central charges, although unconventional in two-dimensional conformal field theories, can arise naturally in both two- and four-dimensional contexts. In two dimensions, they emerge when the Faddeev–Popov ghosts dominate the total central charge, and are closely linked to the resolution of curvature singularities in two-dimensional dilaton gravity theories. Our analysis indicates that this behaviour is not tied to specific models but displays a degree of universality. In four dimensions, conformally invariant scalar fields provide a concrete realization of negative central charges suggesting that the connection between negative central charge and singularity resolution may extend beyond two dimensions.

\section*{ACKWNOWLEDGEMENTS}
A. D. R. acknowledges financial support via  ``{\it Atraccion de Talento Cesar Nombela}'', Grant No. 2023-T1/TEC-29023, funded by Comunidad de Madrid, Spain. F. J. M.-G. is supported by the Ministerio de Ciencia, Innovaci\'on y Universidades, Ph.D. fellowship, Grant No. FPU22/02528. This paper has been supported by Project No. PID2023-149560NB- C21 funded by MCIU /AEI/10.13039/501100011033 / FEDER, UE and by No. CEX2023-001292-S funded by MCIU/AEI.  The paper is based upon work from COST Action CaLISTA CA21109 supported by COST (European Cooperation in Science and Technology).

\end{document}